\documentclass[%
 aip,
 amsmath,amssymb,
 reprint,%
]{revtex4-1}

\usepackage{graphicx}
\usepackage{dcolumn}
\usepackage{bm}

\usepackage[utf8]{inputenc}
\usepackage[T1]{fontenc}
\usepackage{mathptmx}
\usepackage{etoolbox}
\usepackage{hyperref}
\usepackage{physics}
\usepackage{rotating}
\usepackage{multirow}
\usepackage{threeparttable}

\newcommand\PM{\phantom{$-$}}

\makeatletter
\def\@email#1#2{%
 \endgroup
 \patchcmd{\titleblock@produce}
  {\frontmatter@RRAPformat}
  {\frontmatter@RRAPformat{\produce@RRAP{*#1\href{mailto:#2}{#2}}}\frontmatter@RRAPformat}
  {}{}
}%
\makeatother
\begin{document}

\preprint{AIP/123-QED}

\title[]{Benchmarking ionization potentials from pCCD tailored coupled cluster models}
\author{Marta Ga\l{}y\'nska}%
\affiliation{ 
Institute of Physics, Faculty of Physics, Astronomy and Informatics, Nicolaus Copernicus University in Toru\'n, Grudziadzka 5, 87-100 Toru\'n, Poland.
}%
\author{Katharina Boguslawski}%
 \email{k.boguslawski@fizyka.umk.pl}
\affiliation{ 
Institute of Physics, Faculty of Physics, Astronomy and Informatics, Nicolaus Copernicus University in Toru\'n, Grudziadzka 5, 87-100 Toru\'n, Poland.
}

\date{\today}

\begin{abstract}
The ionization potential (IP) is an important parameter providing essential insights into the reactivity of chemical systems.
IPs are also crucial for designing, optimizing, and understanding the functionality of modern technological devices. 
We recently showed that limiting the CC ansatz to the seniority-zero sector proves insufficient in predicting reliable and accurate ionization potentials within an IP equation-of-motion coupled-cluster formalism.
Specifically, the absence of dynamic correlation in the seniority-zero pair coupled cluster doubles (pCCD) model led to unacceptably significant errors of approximately 1.5 eV.
In this work, we aim to explore the impact of dynamical correlation and the choice of the molecular orbital basis (canonical vs.~localized) in CC-type methods targeting 201 ionized states in 41 molecules.
We focus on pCCD-based approaches as well as the conventional IP-EOM-CCD and IP-EOM-CCSD.
Their performance is compared to the CCSDT equivalent and experimental reference data.
Our statistical analysis reveals that all investigated frozen-pair coupled cluster methods exhibit similar performance, with differences in errors typically within chemical accuracy (1 kcal/mol or 0.05 eV).
Notably, the effect of the molecular orbital basis, such as canonical Hartree-Fock or natural pCCD-optimized orbitals, on the IPs is marginal if dynamical correlation is accounted for.
Our study suggests that triple excitations are crucial in achieving chemical accuracy in IPs when modeling electron detachment processes with pCCD-based methods.
\end{abstract}

\maketitle

The reliable determination of ionization potentials (IP) is crucial for the theoretical modeling of molecular electronic structures and molecular properties.
The IP provides information about the system's reactivity as it facilitates measuring the strength of one electron being attached to the molecular bulk and quantifying the molecule's ability to form a more positively charged ion.
This information can be further utilized to design, optimize, and comprehensively understand the functionality of modern technological devices such as photovoltaic (PV) cells, light-emitting diodes, or sensors.\cite{joule-15-precent-osc,opv-19-percent}
For instance, a critical factor in designing novel organic-based donor and acceptor molecules in organic PV devices~\cite{joule-15-precent-osc} is the knowledge of the energies of the highest occupied molecular orbital (HOMO) and the lowest unoccupied molecular orbital (LUMO) and the corresponding HOMO--LUMO gap.
From a theoretical perspective, one of the simplest approximations to deduce orbital energies exploits the diagonal elements of the Fock matrix and the electron repulsion energy.~\cite{piotrus-orbital-energies}
More reliable orbital energies can be, for instance, obtained from the ionization potential (IP)~\cite{Nooijen1992-ip, Nooijen1993-ip, Stanton1994-ip, Stanton1999-ip} and electron affinity (EA)~\cite{musial2014equation} variants of equation-of-motion (EOM)~\cite{rowe1968equations, stanton1993equation,eom-cc-bartlett2012} applied on top of a coupled cluster reference wave function.
The resulting IPs and EAs are then exploited to predict the so-called charge gap.

Apart from CC approaches,~\cite{paldus1992coupled,bartlett2007coupled,monika_mrcc} various calculation protocols have been proposed to investigate the electron detachment process, including density functional theory and its time-dependent formulation,~\cite{yang_parr_book,dft_rev_2012} configuration interaction models,~\cite{foster1960canonical,szalay2012} perturbation theory,~\cite{shavitt_book} algebraic-diagrammatic construction (ADC) schemes,~\cite{Schirmer1998adc,Dempwolff2019adc,Banerjee2019adc,Banerjee2021adc} and Monte Carlo methods.~\cite{qmc,qmc-book-chapter}
These methods feature a diverse spectrum related to their accuracy and computational complexity, cost, and resource requirements.
Furthermore, they can be applied to a wide range of chemical compounds, varying in complexity and size.
Among these methods, different variants of IP-EOM-CC~\cite{nooijen1992coupled,nooijen1993coupled,Stanton1994-ip,Stanton1999-ip,gomes2018ip} have become well-established correlation-based methods, mainly employed for simulating photoelectron spectroscopy.~\cite{geertsen1989equation,stanton1993equation,watts1994inclusion,van2000benchmark,musial2003equation,krylov2008equation,manohar2008noniterative,cooper2010benchmark,evangelista2011alternative,Kus2011,musial2014equation,lischka2018multireference,gulania2021equation,marie2021variational}

The most common CC ansatz is constrained to single and double excitations (CCSD) but can be rather easily extended to perturbatively account for triple substitutions (CCSD(T)), which is commonly known as the gold standard of quantum chemistry.
Those, as well as further extensions of the CC ans{a}tz, including full triples (CCSDT), quadruples, and higher excitations, can be combined with an IP-EOM formalism to describe ionized states.~\cite{musial2003equation} 
Although those EOM-CC methods~\cite{krylov2008equation,ee-cc-methods2012,eom-cc-bartlett2012} are highly reliable in terms of accuracy, they are remarkably expensive and hence limited to relatively small system sizes. 
Thus, a significant effort has been made to devise alternatives of similar accuracy but more reasonable computational complexity.  
The simplified IP-EOM pair coupled cluster doubles (IP-EOM-pCCD) variant~\cite{boguslawski2021open} proved an inexpensive alternative to model open-shell electronic structures within the pCCD~\cite{limacher2013new,boguslawski2014efficient,stein2014seniority} model.
pCCD was originally introduced as a geminal-based wavefunction~\cite{tecmer2022geminal} ansatz using two-electron functions as building blocks for the electronic wave function.~\cite{hurley1953molecular,parr1956generalized,bardeen1957theory,parks1958theory,coleman1965structure,miller1968electron,surjan1999introduction,surjan2012strongly, tecmer2014assessing,johnson2013size, johnson2017strategies,fecteau2020reduced, johnson2020richardson, johnson2022bivariational,faribault2022reduced,fecteau2022near, moisset2022density,tecmer2022geminal} 
Other examples are strictly localized geminals,~\cite{surjan1984interaction,poirier1987application,surjan1994interaction} the antisymmetrized product of strongly orthogonal geminals,~\cite{hurley1953molecular,parks1958theory,kutzelnigg1964direct,surjan1999introduction} and the generalized valence bond perfect pairing~\cite{bobrowicz1977self,cullen1996generalized} models, to name a few.
Such models are a promising alternative to conventional electronic structure approaches, which are typically constructed from one-electron functions.

Initial numerical studies~\cite{boguslawski2021open} demonstrated that the accuracy of IP-EOM-pCCD approaches closely matches the accuracy of CCSD(T) or the density matrix renormalization group~\cite{white,ors_springer,marti2010b,chanreview} (DMRG) algorithm in open-shell electronic structures.
We recently presented a benchmark study to assess the accuracy of the IP-EOM-pCCD method in predicting ionization energies.~\cite{mamache2023}
In Ref.~\citenum{mamache2023}, we compared the vertical ionization energies obtained in the space of one-hole ({1h}) and two-hole-one-particle ({2h1p}) states for three types of molecular orbitals (canonical Hartree-Fock, Pipek-Mezey localized, and natural pCCD orbitals).
Our study suggests that the orbital-optimized IP-EOM-pCCD method, restricted to the {2h1p} operator, demonstrated the highest accuracy among the investigated methods.
However, due to the absence of dynamic correlation, we observed unacceptably large errors in IPs of approximately 1.5 eV.

As demonstrated previously,~\cite{tecmer2023jpcl,leszczyk2021assessing} natural pCCD orbitals present a promising alternative to canonical Hartree-Fock orbitals, serving as a reference wave function for more sophisticated calculations.
Consequently, the question arises whether it is possible to achieve an accuracy comparable to more elaborate approaches (like CCSDT or higher) employing natural pCCD-optimized orbitals in combination with simplified CC ans\"{a}tze, which account for dynamical correlation.
Thus, in the current work, we explore, for the first time, both the effect of dynamical correlation and the choice of the molecular orbital basis (canonical vs.~localized) in CC-type methods including up to double excitations.
Specifically, we focus on various pCCD-tailored CC flavours~\cite{henderson2014seniority,boguslawski2015linearized,leszczyk2021assessing} and compare their performance to the conventional CCSD method exploiting a canonical and pCCD-optimized localized molecular orbital basis.
Specifically, we investigate the influence of dynamical correlation to the IP values determined by six different approaches using the natural pCCD-optimized orbitals, namely frozen-pair (fp)CC~\cite{henderson2014seniority} methods (IP-EOM-fpCCD and IP-EOM-fpCCSD), their linearized (fpLCC)~\cite{boguslawski2015linearized} variants (IP-EOM-fpLCCD and IP-EOM-fpLCCSD), and conventional IP-EOM-CCD and IP-EOM-CCSD~\cite{Nooijen1992-ip, Nooijen1993-ip, Stanton1994-ip, Stanton1999-ip}, and compare their performance to the CCSDT equivalent and experimental reference data.

This work is structured as follows: In section \ref{sec:theory}, we briefly review the investigated theoretical models.
Section \ref{sec:comp} provides an overview of the computational methodology.
Section \ref{sec:results} presents the numerical results, including a statistical analysis.
Finally, we conclude in Section \ref{sec:conclusion}.

\section{THEORY}\label{sec:theory}
The pCCD~\cite{limacher2013new,boguslawski2014efficient,stein2014seniority,tecmer2022geminal} ansatz is a simple reduction of the single-reference CCD approach, where the cluster operator only contains electron-pair excitations $\hat{T}_\textrm{pCCD}$, 
\begin{equation}
\ket{\textrm{pCCD}} = e^{\hat{T}_\textrm{pCCD}} \ket{\Phi_0},
\end{equation}
and
\begin{equation}
\hat{T}_\textrm{pCCD} = \sum_{i}^{n_{\rm occ}} \sum_{a}^{n_{\rm virt}} c_i^a{a_a^\dag}{a_{\bar a}^\dag}{a_{\bar i}}{a_i},
\end{equation}
where $\big|\Phi_0\rangle$ is some reference determinant, $\hat{a}_p$ ($\hat{a}_p^\dag$) are the elementary annihilation (creation) operators for $\alpha$ ($p$) and $\beta$ $(\overline{p})$ electrons, and $c^a_i$ are the pCCD cluster amplitudes.
The above sum runs over all occupied $i$ and virtual $a$ orbitals.
Typically, the pCCD molecular orbitals are optimized,~\cite{boguslawski2014nonvariational,boguslawski2014projected,boguslawski2014efficient,stein2014seniority} which re-establishes size consistency and yields localized and symmetry-broken orbitals that allow us to simulate quantum states with (quasi-)degeneracies.~\cite{boguslawski2016analysis}
Numerical examples comprise bond-breaking processes in small molecules~\cite{tecmer2014assessing,limacher2015orbital,tecmer2015singlet,brzek2019benchmarking,henderson2019geminal,nowak2021orbital,leszczyk2021assessing,leszczyk2022}, heavy-element-containing compounds featuring lanthanide~\cite{tecmer2019modeling} or actinide~\cite{tecmer2015singlet,garza2015actinide,boguslawski2016targeting,boguslawski2017erratum,nowak2019assessing,leszczyk2022,Nowak2023,chakraborty2023} atoms, organic electronics\cite{jahani2023relationship,tecmer2023jpcl}, and electronically excited states.~\cite{boguslawski2016targeting,boguslawski2017erratum,boguslawski2018targeting,nowak2019assessing,kossoski2021excited,boguslawski2021open,bartlett-pccd-tcc}
Although these numerical studies support pCCD to be a promising alternative
to capture static/nondynamic electron correlation effects,~\cite{sinanoglu1963, bartlett_1994,  entanglement_letter}
a large fraction of the correlation energy cannot be captured by electron-pair states alone.
This missing correlation energy is commonly attributed to so-called broken-pair states.
These correlation effects are commonly included \textit{a posteriori} using various state-of-the-art techniques.~\cite{tecmer2022geminal}

One possibility to account for dynamical correlation is to exploit a coupled cluster correction with a pCCD reference function.~\cite{henderson2014seniority,boguslawski2015linearized,leszczyk2021assessing}
For instance, in the frozen-pair coupled cluster (fpCC) ansatz,~\cite{henderson2014seniority,leszczyk2021assessing} 
\begin{equation}
\ket{{\rm fpCC}} = e^{\hat{T}^{\rm ext}} \ket{{\rm pCCD}} = e^{\hat{T}^{\rm ext}} e^{\hat{T}_{\rm pCCD}} \ket{\Phi_0},
\end{equation}
the pCCD wave function is taken as the fixed reference function and the $\hat{T}^{\rm ext}$ cluster operator contains electron excitations beyond electron-pair excitations.
Thus, the cluster operator of fpCCD is defined as $\hat{T}^{\rm ext} = \hat{T}_2^\prime = \hat{T}_2 - \hat{T}_{\rm pCCD}$, while the cluster operator of fpCCSD includes also single excitations, $\hat{T}^{\rm ext} = \hat{T}_1 + \hat{T}_2^\prime$.
We should stress that fpCC theory can be considered as a conventional tailored coupled cluster approach.~\cite{kinoshita2005, tailoredcc2006, lyakh2011, tailoredcc2012}
The fpCC ansatz can be further simplified by truncating the
Baker–Campbell–Hausdorff expansion after the second term (concerning all non-pair excitations) and hence including only linear terms in $\hat{T}^{\rm ext}$.~\cite{boguslawski2015linearized,boguslawski2017benchmark}
Strictly speaking, the frozen-pair Linearized CC correction (fpLCC) does not fall into the category of tailored CC methods.
Nonetheless, we will use the acronym fpLCC due to its simplicity (originally, fpLCC was introduced as pCCD-LCC).
The wave function ansatz of fpLCC is approximated as
\begin{equation}\label{eq:lcc}
\ket{{\rm fpLCC}} \approx (1+\hat{T}^{\rm ext})  \ket{{\rm pCCD}} = (1+\hat{T}^{\rm ext}) e^{\hat{T}_{\rm pCCD}} \ket{\Phi_0}.
\end{equation}
We should note that all disconnected terms containing $\hat{T}_{\rm pCCD}$ still appear in the fpLCC amplitude equations as the exponential ansatz of pCCD is not linearized.
For instance, terms associated with $\hat{T}^{\rm ext}\hat{T}_{\rm pCCD}$ and $\hat{T}_{\rm pCCD}^2$ have to be considered in fpLCCD-type methods.
Thus, in fpLCC, the coupled cluster equations are linear concerning non-pair amplitudes
$\hat{T}^{\rm ext}$ but the coupling between all pair- and non-pair amplitudes is included.

\begin{figure}[t]
\centering
  \includegraphics[width=80.0mm]{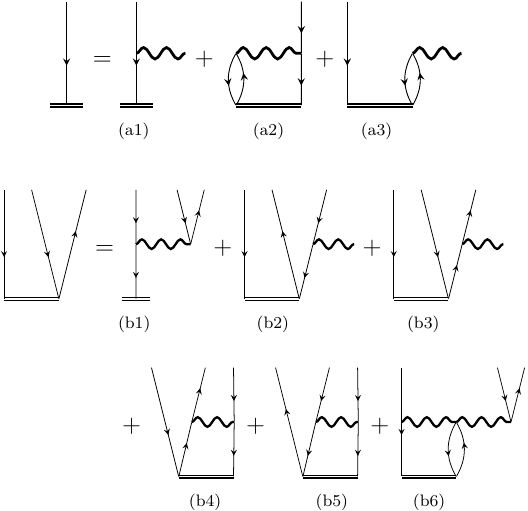}
  \caption{Diagrammatic representation of the IP-EOM-fp(L)CCSD equations (antisymmetrized formalism).} 
  \label{fgr:diag}
\end{figure}

Since we formally work in a single-reference CC picture, we can straightforwardly employ single-reference CC techniques to target, for instance, electronically excited, ionized, and electron-attached states.~\cite{musial2014equation,perera2016singlet,gulania2021equation}
Another possible extension are spin-flip EOM-CC methods.~\cite{Casanova2020}
Specifically, for ionized states, we can employ the IP-EOM formalism, where we use a linear ansatz on top of the closed-shell fp(L)CC reference function to parametrize the k-th (ionized) state
\begin{equation}
\ket{\Psi_k} = \hat R(k)\ket{\textrm{fpCC}}
\end{equation}
where the operator $\hat R(k)$ generates the targeted ionized state $k$ from the initial fp(L)CC reference state.
In the single IP-EOM formalism, $\hat R(k)$ reads
\begin{equation}
\hat R^{\rm IP} = \sum_{i}r_i\hat a_i+ \frac{1}{2}\sum_{ij{a}}{r}^a_{ij}\hat{a}_a^\dag \hat a_j \hat a_i+ \dots ~ = \hat R_{1h}+\hat R_{2h1p}+\dots ~ 
\end{equation}
where we introduced the hole (h, encoding $\hat a_i$) and particle (p, encoding $\hat a ^\dagger_a$) labels and dropped the k-dependence for better readability.
The ionized states are then obtained by solving the corresponding EOM equations
\begin{equation}
{[\hat{H}_N,\hat R ]} \ket{\textrm{fpCC}} =  \omega \hat R \ket{\textrm{fpCC}},
\end{equation}
where $ \omega = \Delta E - \Delta E_0$ is the energy corresponding to the ionization process concerning the fpCC ground state,
while $\hat{H}_N = \hat{H} - \langle\Phi_0\big|\hat{H}|\Phi_0\rangle  $ is the normal product form of the Hamiltonian.
We can rewrite the above equation in the well-known form
\begin{equation}
{\cal H}^\textrm{fpCC}_N \hat R\ket{\Phi_0} =  \omega \hat R \ket{\Phi_0}
\end{equation}
where ${\cal H}^\textrm{fpCC}_N$ is the similarity transformed Hamiltonian of the used fpCC flavour in its normal-product form
\begin{equation}
    {\cal H}^\textrm{fpCC}_N= e^{- \hat{T}_\textrm{pCCD}}e^{- \hat{T}^\textrm{ext}}  \hat{H}_N e^{\hat{T}^\textrm{ext}}e^{\hat{T}_\textrm{pCCD}}.
\end{equation}
The ionization energies are obtained by iteratively diagonalizing ${\cal H}^\textrm{fpCC}_N$ of the chosen pCCD-based CC correction.
Note that depending on the selected CC model, the similarity transformed Hamiltonian either contains all non-vanishing non-linear terms (here, ${\cal H}^\textrm{fpCCD}_N$ and ${\cal H}^\textrm{fpCCSD}_N$) or only non-linear terms associated with the electron-pair excitation operator (here, ${\cal H}^\textrm{fpLCCD}_N$ and ${\cal H}^\textrm{fpLCCSD}_N$.
The diagrammatic representation of the IP-EOM-fpCC equations is shown in Fig.~\ref{fgr:diag} (see also, for instance, Ref.~\citenum{musial2003equation} for the diagrammatic form and its algebraic expressions).
We should note that the IP-EOM-fpCC(S)D equations are similar in form to the conventional IP-EOM-CCSD formalism.
However, in the IP-EOM-fpLCCSD method, the effective Hamiltonian diagrams (a1), (b1), (b2), (b3), (b4), (b5) do not contain the disconnected $\hat{T}_1\hat{T}_1$ terms, while, in addition, diagram (b1) lacks the $\hat{T}_1\hat{T}_1\hat{T}_1$ part.
Furthermore, the $\hat{T}_1\hat{T}_2$ term contained in (b1) is replaced by the simpler $\hat{T}_1\hat{T}_p$ counterpart due to the truncation of the BCH expansion of the LCC correction (see also eq.~\eqref{eq:lcc}).
Finally, we should note that we focused on the $S_z=-0.5$ case~\cite{musial2014equation,boguslawski2021open} in its spin-dependent and spin-summed versions.
The spin-summed working equations can be obtained by either diagrammatic or algebraic spin summation (see also Ref.~\citenum{musial2003equation}) of the IP-EOM-fp(L)CC equations shown in Fig.~\ref{fgr:diag}.
For the latter case, the targeted ionized (open-shell) states are the doublet states of the $S_z=-0.5$ spin block.

\begin{figure}[t]
\centering
  \includegraphics[width=90.0mm]{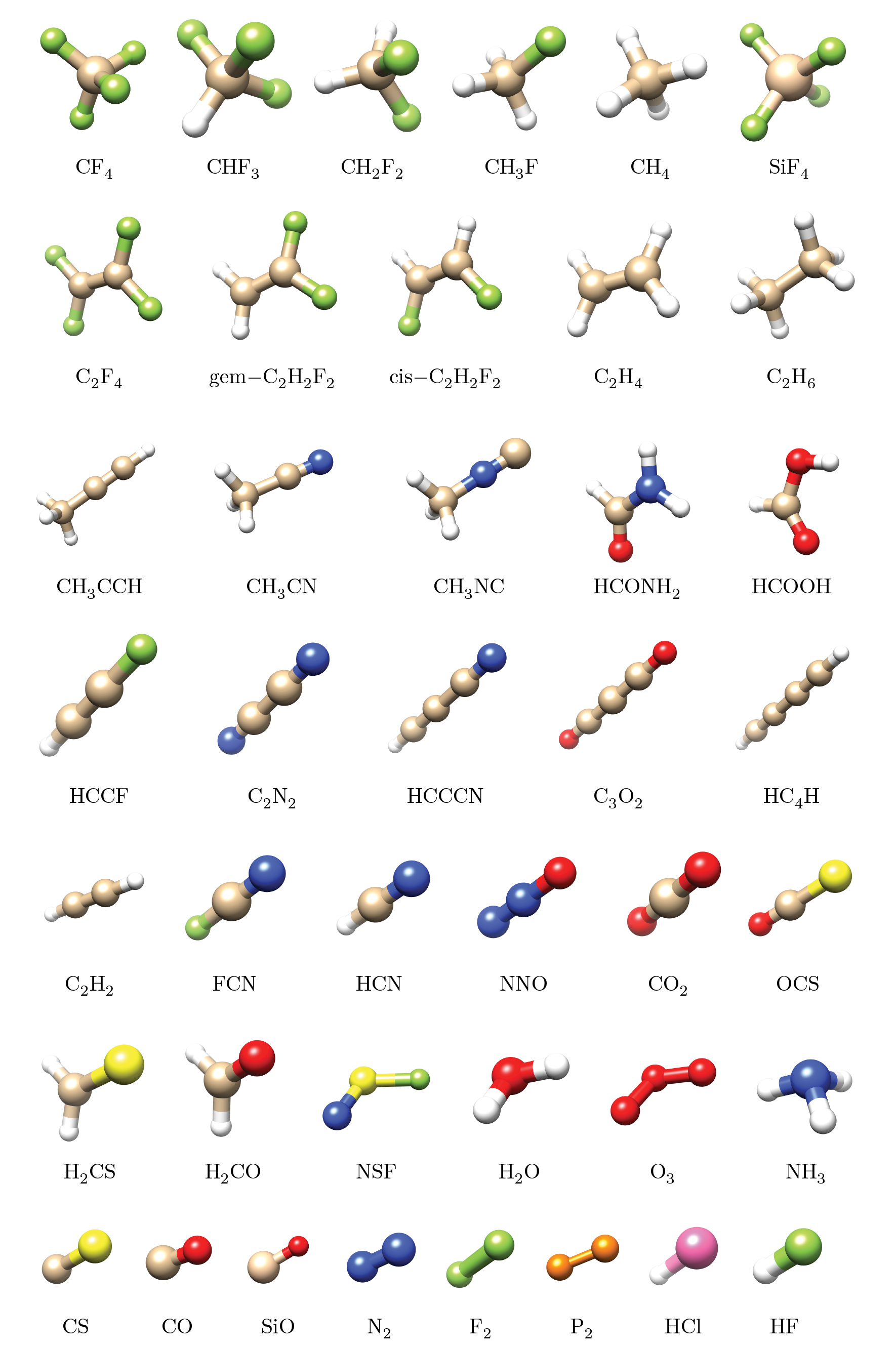}
  \caption{The benchmark set containing 41 molecules relaxed at the CCSD(T)/aug-cc-pVTZ level provided in Ref.~\citenum{ranasinghe2019vertical}.} 
  \label{fgr:molecules}
\end{figure}

\begin{table}[b!]
\caption{Statistical error measures in electronvolts (eV), such as mean error (ME), mean absolute error (MAE), and root-mean-square error (RMSE), were assessed based on the ionization potentials (IP) calculated using various methods: IP-EOM-fpCCD, IP-EOM-fpCCSD, IP-EOM-fpLCCD, IP-EOM-fpLCCSD, IP-EOM-CCD(pCCD), and IP-EOM-CCSD(pCCD), listed in the upper part of each block, are derived in this work. All these CC approaches exploit natural pCCD-optimized orbitals (or the orbital-optimized pCCD reference determinant) and are conducted in the space of two-hole-one-particle ({2h1p}) states. The corresponding errors in IPs for CCSD done with Hartree-Fock orbitals, unitary coupled cluster (UCC), algebraic-diagrammatic construction (ADC) methods, and pCCD, using natural pCCD-optimized orbitals, are presented in the lower part of each block. The errors in IPs were calculated concerning IP-EOM-CCSDT~\cite{ranasinghe2019vertical} (top) and experimental data~\cite{ranasinghe2019vertical} (bottom). The definitions for ME, MAE, and RMSE are printed in the table footnote.}
\label{tab:errors}
\begin{tabular*}{0.48\textwidth}{@{\extracolsep{\fill}}lccc}
\hline
 & \multicolumn{3}{c}{Errors w.r.t.~IP-EOM-CCSDT} \\\cline{2-4}
Method & ME & MAE &  RMSE\\
\hline
 IP-fpCCD                    & \PM0.295	& 0.346 & 0.461 \\
 IP-fpLCCD                   & \PM0.335	& 0.382	& 0.489 \\
 IP-CCD(pCCD)                & \PM0.241	& 0.297	& 0.418\\
 IP-fpCCSD                   & \PM0.239	& 0.293	& 0.399 \\
 IP-fpLCCSD                  & \PM0.293	& 0.341 & 0.432\\
 IP-CCSD(pCCD)               & \PM0.183	& 0.238	& 0.350 \\
 \hline
IP-pCCD\cite{mamache2023}                         & $-$1.535  & 1.535 & 1.633\\
IP-CCSD(HF)\cite{ranasinghe2019vertical}         & \PM0.186 & 0.241 & 0.341\\
IP-UCC2\cite{dempwolff2022vertical}              & $-$0.488 & 0.579 & 0.728\\
IP-UCC3\cite{dempwolff2022vertical}              & \PM0.260 & 0.306 & 0.377\\
IP-ADC(2)\cite{dempwolff2022vertical}            & $-$0.545 & 0.607 & 0.737\\
IP-ADC(3(3))\cite{dempwolff2022vertical}         & \PM0.269 & 0.351 & 0.442\\
IP-ADC(3(4+))\cite{dempwolff2022vertical}        & \PM0.306 & 0.339 & 0.418\\
IP-ADC(3(DEM))\cite{dempwolff2022vertical}       & \PM0.292 & 0.334 & 0.411\\
\hline
 & \multicolumn{3}{c}{Errors w.r.t.~experiment} \\\cline{2-4}
Method & ME & MAE &  RMSE\\
\hline
 IP-fpCCD                    & \PM0.259  & 0.336 & 0.457 \\
 IP-fpLCCD                   & \PM0.299  & 0.361 & 0.481 \\
 IP-CCD(pCCD)                & \PM0.206  & 0.303 & 0.422 \\
 IP-fpCCSD                   & \PM0.203  & 0.283 & 0.391 \\
 IP-fpLCCSD                  & \PM0.256  & 0.311 & 0.415 \\
 IP-CCSD(pCCD)               & \PM0.147  & 0.253 & 0.253 \\
 \hline
IP-pCCD\cite{mamache2023}                       & $-$1.570 & 1.570 & 1.697\\ 
IP-CCSD(HF)\cite{ranasinghe2019vertical}        & \PM0.150 & 0.252 & 0.351\\
IP-CCSDT\cite{ranasinghe2019vertical}           & $-$0.035 & 0.197 & 0.262\\
IP-UCC2\cite{dempwolff2022vertical}             & $-$0.523 & 0.684 & 0.841\\
IP-UCC3\cite{dempwolff2022vertical}             & \PM0.225 & 0.312 & 0.394\\
IP-ADC(2)\cite{dempwolff2022vertical}           & $-$0.580 & 0.693 & 0.844\\
IP-ADC(3(3))\cite{dempwolff2022vertical}        & \PM0.233 & 0.356 & 0.419\\
IP-ADC(3(4+))\cite{dempwolff2022vertical}       & \PM0.271 & 0.342 & 0.423\\
IP-ADC(3(DEM))\cite{dempwolff2022vertical}      & \PM0.256 & 0.339 & 0.419\\
\hline
\end{tabular*}
\begin{tablenotes}
   \item[*] ME = $\sum_i^N \frac{E_i^{\rm method} - E_i^{\rm ref}}{N} $
   \item[*] MAE = $\sum_i^N \frac{|E_i^{\rm method} - E_i^{\rm ref}|}{N} $
   \item[*] RMSE = $\sqrt{\sum_i^N \frac{(E_i^{\rm method} - E_i^{\rm ref})^2}{N}}$
\end{tablenotes}
\end{table}

\section{COMPUTATIONAL DETAILS}\label{sec:comp}
The vertical ionization potentials (IP) were calculated using different CCD- and CCSD-type flavours as implemented in a developer version of the PyBEST v1.4.0-dev0 software package.~\cite{boguslawski2021pythonic,boguslawski2024pybest, pybest_zendo}
These included IP-EOM-fpCCD, IP-EOM-fpCCSD, IP-EOM-fpLCCD, IP-EOM-fpLCCSD, IP-EOM-CCD, and IP-EOM-CCSD.
In all CC calculations (electronic ground states and IP-EOM), the pCCD-optimized orbitals (labeled as ``(pCCD)'') were used to construct the reference determinant $\ket{\Phi_0}$, which were obtained through a variational orbital-optimization protocol of the pCCD reference calculation.~\cite{boguslawski2014efficient,stein2014seniority,boguslawski2014nonvariational, boguslawski2014projected}
The optimization protocol used in pCCD calculations automatically selects the reference determinant according to the pCCD natural occupation numbers.

The ionization energies were computed in the space of two-hole-one-particle ({2h1p}) states as previous studies indicate that errors in IPs are significantly reduced compared to the use of only one-hole ({1h}) states.~\cite{mamache2023}
No symmetry constraints were applied to allow the algorithm to freely relax the orbitals resulting in a symmetry-broken, localized molecular orbital basis.

A frozen core was used in all CC calculations, keeping the 1s orbitals for C, N, O, and F and 1s, 2s, and 2p orbitals for Si, P, S, and Cl frozen.
We should note that preliminary tests indicate minimal impact on the IP values when freezing core orbitals.
All calculations employed the cc-pVTZ basis set by Dunning,~\cite{dunning1989gaussian} facilitating a direct comparison with previously published vertical IPs obtained using the IP-EOM-CCSDT model,~\cite{ranasinghe2019vertical} various forms of the unitary coupled-cluster (IP-UCC) approach, and algebraic-diagrammatic construction (IP-ADC) methods.\cite{dempwolff2022vertical} 

Furthermore, our test set contains 41 molecules shown in Fig.~\ref{fgr:molecules}, whose molecular geometries were relaxed using the CCSD(T)/aug-cc-pVTZ method~\cite{kendall1992electron} and are available in the supplementary material of Ref.~\citenum{ranasinghe2019vertical}.
In total, we optimized 201 IP states and compared the performance of our IP-EOM-CC methdos to theoretical and experimental reference data.

\section{RESULTS AND DISCUSSION}\label{sec:results}
\begin{figure*}
\centering
\includegraphics[width=0.9\textwidth]{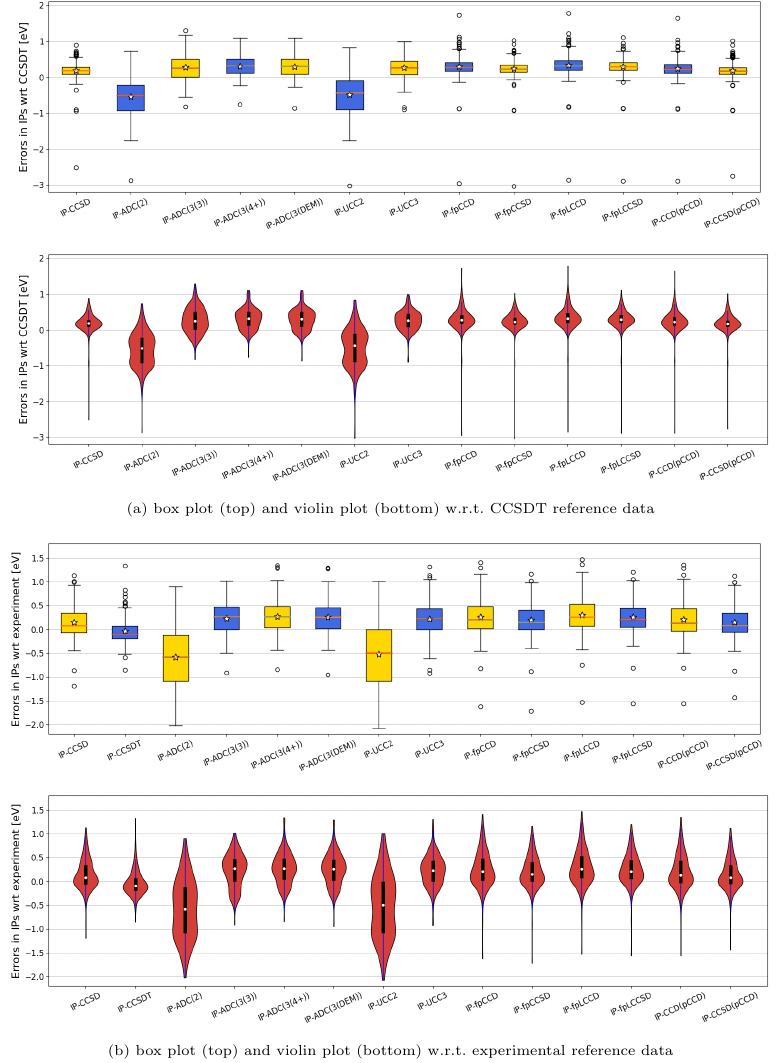}
\caption{Box plots presented at the top and violin plots at the bottom, illustrating errors [eV] derived from selected methods (refer to Table 1 for numerical values). All errors are reported relative to either (a) IP-EOM-CCSDT or (b) experimental reference data. For brevity, we have omitted the EOM prefix in IP-EOM-CC-type methods. A star in each box plot denotes the mean value, while a white dot in each violin plot represents the median value.}
\label{fig:errorplots}
\end{figure*}

Table~\ref{tab:errors} summarizes our statistical analysis, including mean errors (ME), mean absolute errors (MAE), and root-mean-square errors (RMSE) calculated using IP-EOM-fpCCD, IP-EOM-fpCCSD, IP-EOM-fpLCCD, IP-EOM-fpLCCSD, conventional IP-EOM-CCD(pCCD), and IP-EOM-CCSD(pCCD) of 201 ionized states in 41 molecules shown in Fig~\ref{fgr:molecules}.
The footnote in Table~\ref{tab:errors} defines the error measures used in this work.
We should note that we use the labels CCD(pCCD) and CCSD(pCCD) to indicate that the corresponding CCD and CCSD calculations were done employing pCCD-optimized natural orbitals (or equivalently the orbital-optimized pCCD reference determinant in the CC ansatz).
The upper section of Table~\ref{tab:errors} presents error values concerning IP-EOM-CCSDT reference data,~\cite{ranasinghe2019vertical} while the lower section reports the corresponding errors relative to experimental results.\cite{ranasinghe2019vertical}
For a direct comparison, the table includes data from IP-EOM-CCSD calculated with a canonical Hartree-Fock reference determinant (or molecular orbital basis) indicated as IP-CCSD(HF), two variations of the unitary coupled cluster ans{a}tz (IP-UCC2 and IP-UCC3), and four variants of non-Dyson algebraic diagrammatic construction schemes (ADC(2), ADC(3(3)), ADC(3(4+)), and ADC(3(DEM)).~\cite{dempwolff2022vertical}
Furthermore, Fig.~\ref{fig:errorplots}(a) visualizes, using box and violin plots, the locality, spread, skewness, and distribution of errors of all targeted ionization potentials concerning IP-EOM-CCSDT.
Fig.~\ref{fig:errorplots}(b) displays an equivalent analysis for experimental data.
The individual ionization energies obtained by all methods investigated in this work are accessible in the Electronic Supplementary Information (ESI)$^\dag$.

The effect of adding dynamical correlation, that is including the seniority two and seniority four sectors in the fpCC reference function of the IP-EOM-fpCCD method, provides a considerable improvement of the ionization energies in comparison to the seniority zero orbital-optimized pCCD method.
Specifically, the MAE and RMSE errors are reduced by approximately 1.2 eV on average if we go beyond the seniority-zero sector in the CC reference function.
This supports the original finding~\cite{mamache2023} that dynamical correlation is needed to correctly describe the electron detachment process within pCCD-based methods.
Using a simplified version of the frozen-pair methods, IP-EOM-fpLCCD increases slightly the ME, MAE, and RSME errors compared to IP-EOM-fpCCD by 0.040 (0.040) eV, 0.036 (0.025) eV, 0.028 (0.024) eV, respectively, with respect to IP-EOM-CCSDT reference data (experiment).
On the other hand, IP-EOM-CCD(pCCD) (conventional IP-EOM-CCD with an orbital-optimized pCCD reference determinant) decreases the errors of IP-EOM-fpCCD by 0.054 (0.053) eV for ME, 0.049 (0.033) eV for MEA, and 0.043 (0.035) eV for RMSE with respect to IP-EOM-CCSDT (experimental) results.  


A similar trend is observed for CCSD-based approaches.
Specifically, IP-EOM-fpLCCSD yields the largest errors, while IP-EOM-CCSD(pCCD) (conventional IP-EOM-CCSD with an orbital-optimized pCCD reference determinant) exhibits the smallest errors compared to theoretical (experimental) reference values.
Nonetheless, their differences in errors are acceptable, amounting up to around 0.11 eV (or 2.5 kcal/mol).
Most importantly, including single excitations slightly reduces the ME, MAE, and RMSE values by 0.041 to 0.061 eV between the frozen-pair CCD and CCSD methods, respectively.
Regarding statistical errors, IP-EOM-fpCCSD is the most accurate among all investigated pCCD-based approaches investigated in this work.
Specifically, it yields very similar errors to the IP-EOM-UCC3 method identified as the best among recently investigated approximations.~\cite{ranasinghe2019vertical}
Concerning IP-EOM-CCSDT (experiment), the relative error measures between IP-EOM-fpCCSD and IP-EOM-UCC3 are $\Delta\Delta$ME = 0.021 (0.022) eV, $\Delta\Delta$MAE = 0.013 (0.029) eV, and $\Delta\Delta$RMSE = $-$0.022 (0.003) eV, where $\Delta\Delta$ indicates the difference between the IP-EOM-fpCCSD and IP-EOM-UCC3 error measures.

Our box plots (see Fig.~\ref{fig:errorplots} top) illustrate that the differences in errors among all fpCC methods are very similar, displaying an almost identical dispersion of 50\% of errors (highlighted in yellow and blue boxes).
The total range of scope (indicated by the whiskers) diminishes slightly with the addition of single excitations.
However, the differences are minimal.
The violin plots (see Fig.~\ref{fig:errorplots} bottom) highlight interquartile ranges distributed closely around the median.
The skewness of errors is left-shifted in all cases.
Although the dispersion of results is slightly smaller when using the IP-EOM-CCSDT method as a reference, a similar trend of error dispersion can be seen for both references (theoretical and experimental).
All frozen-pair variants exhibited an accuracy range similar to IP-EOM-CCSD conducted with canonical HF molecular orbitals, demonstrating a comparable dispersion and skewness of errors.

The IP-EOM-CCSD results reported by Ranasinghe et al.\cite{ranasinghe2019vertical} allow us to directly assess the effect of the choice of the molecular orbital basis on molecular properties, that is if the performance is significantly different between canonical HF and natural pCCD-optimized orbitals.
Surprisingly, the errors are almost identical and exhibit resilience to the choice of the reference wave function.
Furthermore, the overall appearance of error distribution, skewness, and even the positioning of outliers presented in the box and violin plots in Fig.~\ref{fig:errorplots} is almost identical in both cases (differences lie within chemical accuracy).
This suggests that including triple excitations in the theoretical model will be crucial for further improving the accuracy of pCCD-based approaches in predicting IPs.

\section{CONCLUSIONS}\label{sec:conclusion}
As recently shown,~\cite{mamache2023} restricting the CC ansatz to the seniority-zero sector is insufficient in predicting reliable and accurate IPs.
Although the seniority-zero pCCD model can capture static correlation reliably, it is inadequate to describe electron detachment with sufficient accuracy.
This deficiency was attributed to the missing broken-pair states, that is the exclusion of the seniory-two, seniority-four, etc.~sectors.
In this work, we investigated the impact of dynamical correlation and the choice of the molecular orbital basis (canonical vs.~localized) on vertical ionization potentials using various pCCD-based and conventional CC approaches.
Specifically, we studied six CC variants: IP-EOM-fpCCD, IP-EOM-fpLCCD, IP-EOM-CCD(pCCD), IP-EOM-fpCCSD, IP-EOM-fpLCCSD, and IP-EOM-CCSD(pCCD).
Throughout this work, we included (up to) 2p1h operators in the IP-EOM formalism as the resulting pCCD-based model turned out to be superior to the corresponding IP-EOM approach restricted to 1h operators.~\cite{mamache2023}
Our analysis encompasses a set of 41 molecules, targeting 201 ionized states.
These ionization energies are compared to IP-EOM-CCSDT and experimental reference data. Furthermore, our results are juxtaposed with those obtained using various conventional CC methods, UCC flavors, and non-Dyson ADC second and third-order schemes.

Our statistical analysis (mean errors, mean absolute errors, root mean square errors) highlights that all investigated frozen-pair coupled cluster methods feature similar performance.
Specifically, the differences in errors are typically within chemical accuracy (1 kcal/mol or 0.05 eV).
Adding single excitations slightly reduces error measures with respect to the corresponding CCD model.
Yet, these changes approach chemical accuracy, constituting approximately 0.06 eV.
Our benchmark data renders IP-EOM-fpCCSD the best-performing method among all tested frozen-pair variants.
We should stress, however, that differences between the investigated frozen-pair methods are nearly invisible on box and violin plots.
Furthermore, the scattering of errors and their distribution around the median make them comparable to the conventional IP-EOM-CCSD method.
On the other hand, the error measures of IP-EOM-fpCCSD are comparable to the accuracy of IP-EOM-UCC(3), identified as the best among recently investigated approximations.~\cite{ranasinghe2019vertical}
Finally, the influence of the molecular orbital basis or CC reference determinant (that is canonical vs.~localized) is marginal as the conventional IP-EOM-CCSD and IP-EOM-CCSD(pCCD) result in almost identical errors in ionization potentials.
This observation suggests that triple excitations are crucial for further improving IPs and approaching chemical accuracy for modeling electron detachment processes with pCCD-based methods.

\section*{Acknowledgements}
The research leading to these results has received funding from the Norway Grants 2014--2021 via the National Centre for Research and Development.
M.G.~acknowledges financial support from a Ulam NAWA -- Seal of Excellence research grant (no.~BPN/SEL/2021/1/00005). We acknowledge that the results of this research have been achieved using the DECI resource Bem (Grant No.~412) based in Poland at Wroclaw Centre for Networking and Supercomputing (WCSS, http://wcss.pl) with support from the PRACE aisbl.


\renewcommand\refname{References}

\bibliography{rsc} 
\bibliographystyle{rsc} 

\end{document}